%% file: borrador.tex
\documentclass[10pt,journal]{IEEEtran}

\usepackage[utf8]{inputenc}

\usepackage[T1]{fontenc}

\usepackage{makeidx}
\usepackage{siunitx}

\usepackage{graphicx}
\usepackage{float} 

\usepackage{amsmath}
\usepackage{amsfonts}
\usepackage{amssymb}

\usepackage{physics}

\usepackage{multirow} 
\usepackage{multicol} 
\usepackage[table]{xcolor}

\usepackage[pdftex]{hyperref}
\definecolor{lightgray}{gray}{0.9}

\begin{document}



\title{Variations of the SIR model for COVID-19 evolution }

\author{Nana Cabo Bizet$^a$, Jonathan Hidalgo Nu\~nez$^b$, Gil Estefano Rodr\'{\i}guez Rivera$^c$. \\ \medskip
{\it Laboratorio de Datos, Departamento de F\'{\i}sica, }\\ {\it División de Ciencias e Ingenierías,} \\ {\it Universidad de Guanajuato.}}


\maketitle{}  
\footnote{$^a$ nana@fisica.ugto.mx}
\footnote{$^b$ ja.hidalgonunez@ugto.mx}
\footnote{$^c$ ge.rodriguezrivera@ugto.mx}

\begin{abstract}
    In this work, we discuss the SIR epidemiological model and different variations of it applied to the propagation of the COVID-19 pandemia; we employ the data of the state of Guanajuato and of Mexico.  We present some considerations that can improve the predictions made by those models. We consider a time-dependent infection rate, which we adjust to the data. Starting from a linear regime where the populations are much smaller that the country or state population and the population of susceptible (S) can be approximated in convenient units to $S \sim 1$, we make fits of the parameters. We also consider the case when the susceptible starts departing from 1, for this case we adjust an effective contagion rate. We also explore the ratio of detected populations and the real ones, obtaining that -for the analyzed case it is of $\sim 10 \%$. 
    We estimate the number of deaths by making a fit versus the recovered cases, this fit is in first approximation linear, but other powers can give a good agreement. By predictions to past data, we conclude that adaptations of the SIR model can be of great use in describing pandemia´s propagation, specially in limited time periods. 
    
\end{abstract}

\section{Introduction}

The SIR model is a well-established tool for making predictions about the evolution of an epidemic \cite{lobo,brauer,mingmingli}. In the years of the pandemic a variety of models doing modifications to the SIR model have been considered\cite{tangetal,li, acunazegarra,Moriarty,kumar,palladino,lisashiller,zhao,walker,li,bernaldelepine,gonzalezurena,Hyokyoung,Boatto,palladino,diamondprincess,Bhatia,cheng}. It is the core of many more complex population evolution models, with more compartments. In this work the objective is to make some modifications of the SIR model to make better predictions. Based on previous work we consider a time-dependent infection rate \cite{cabobizet,mayorgacabo}. The interest in using
a simple model as the SIR is because we expect to find  models requiring minimal data to work.
They should reflect the complexities of the real world without making many assumptions.

For example, here it is possible to make an estimation of the detection rate  just with the 
knowledge of the populations of infected, recovered, and deceased cases. By studying such frameworks, it is necessary to fit some subsets of the population, like  the recovered and the deceased as a function of the \textit{immune} cases (the recovered plus the deceased). Although a linear fit is good enough for predictions in a small time window, the y-axis intersection of such fits reveals their problems with them.

In \textbf{Section \ref{sec 2}} we discuss the basic equations of the SIR model. We employ populations in an adimensional notation by defining: $S\to S/N$, $R\to R/N$, $D\to D/N$ and $I\to I/N$. In this section, we introduce the effective recovery rate $\gamma_{eff}$ and the infection rate $\beta$: they are important because the following sections will be built over these quantities. In \textbf{Section \ref{sec 3}} we study briefly how to calculate the real infection rate $\beta_r$. Then, we discuss what happens when $I+D+R<<N$; we use the approximated infection rate $\beta_{approx}$ under those specific conditions. In \textbf{Section \ref{sec 4}} we present the $\beta_S$ model. After a brief discussion of its benefits, it is shown how it transforms between the detected cases and the total ones. Unlike the SIR model and the approximated model, it is not invariant under that transformation. In the previous discussion, we introduce the detection rate $k$, which will be studied profoundly in \textbf{Section \ref{sec 6}}: from its intuitive justification to how to estimate it via the $\beta_S$ model. Turns out that the $\beta_S$ model is very useful to calculate the effective detection rate $k_{eff}$. In \textbf{Section \ref{sec 5}} we show some comparisons between the predictions made using the real transmission rate $\beta_r$ and the ones made using $\beta_S$ at different stages of the pandemic. In \textbf{Section \ref{sec 7}} we discuss a model that takes into consideration directly the vaccination process adding an extra term accounting for the vaccination rate $\Tilde{\beta}$. We also compare this model to the $\beta_S$ model and the conditions in which each works. In \textbf{Section \ref{sec 8}} we show a comparison of the predictions made using the real transmission rate $\beta_r$, the ones made using the $\beta_S$ model and the ones made with the vaccination model. In \textbf{Section \ref{sec 9}} we discuss the consequences of using the immune $\Tilde{R}$ directly in the model instead of using the recovered $R$ and the deceased $D$ each by their own. By itself, it is needed to define a function that relates the directly predicted populations $\Tilde{R}$ and $I$ and the other ones $R$ and $D$. We used a linear fit when we made predictions. But this linear fit leads to some interpretation problems that may reflect or not the problems with the data recollection and reporting process, we discuss those possibilities.

\section{SIR Model and beta in linear regime}\label{sec 2}
The \textbf{SIR model} is one of the simplest models to emulate the evolution infectious diseases and it is described by the following equations:
\begin{align*}
\frac{dS}{dt} &= - \beta S I \\
\frac{dI}{dt} &= \beta S  I - \gamma_{eff} I\\
\frac{d\Tilde{R}}{dt} &= \gamma_{eff} I,
\end{align*}
where
\begin{itemize}
    \item $I$ : Actual number of infected people.
    \item $\Tilde{R}$ : \textit{Immune} people ($R+D$).
    \item $D$ : Deceases due to the pandemic.
    \item $R$ : Recovered people.
    \item $\beta$ : Infection rate.
    \item $\gamma_{eff}$ : Effective recovery rate.
    
\end{itemize}
In the convention employed the populations are adimensional,
and the constants have units $[\gamma_{eff}]=[\beta]=1/time$.

$\gamma_{eff}$ is the effective recovery rate and is calculated by fitting a slope between successive values of $\Tilde{R}(t+1)-\Tilde{R}(t)$ versus $I(t)$, in which the parameter $t$ represents the day of evaluation.
\\
In \textbf{Section \ref{sec 12}} we discuss how SIR equations change working with reported data and adapt them to predict real data.

\section{Estimating the real transmission rate beta}\label{sec 3}
By substituting $S+\Tilde{R}+I=1$ (in the adimensional populations convention) into the second SIR equation, we obtain that the real transmission rate can be calculated with
\begin{equation}
Y_{\beta_r}=\left(\frac{\dot{I} }{I}+\gamma_{eff}\right)\frac{1}{1-(I+\Tilde{R})}
\end{equation}
Where $\dot{I}$ is defined as $I(t+1)-I(t)$ and $1$ is the total (fixed) adimensional population. The real infection rate defined as $\beta r$ is obtained fitting $Y_{\beta_r}$ against the time with an exponential regression.  

The quantity $(I+\Tilde{R})$ is close to $0$ on early stages of the pandemic (because $I+\Tilde{R}<<1$), so we can made the approximation  $\beta_r \approx \beta_{approx}$ for the first months. Where $\beta_{approx}$ is fitted using $Y_{\beta_{approx}}$ instead of $Y_{\beta_{r}}$, where:
\begin{equation}
    Y_{\beta_{approx}}=\frac{\dot{I} }{I}+\gamma_{eff}
\end{equation}

\input{Sections/4betaS}

\section{Comparison of the estimations of the contagion at different betas}\label{sec 5}

In this section, we compare the real $\beta$ and $\beta_S$ and how it changes the predicted values of infected in different periods of time.

\begin{figure}[H]
    \centering
    \includegraphics[scale=.5]{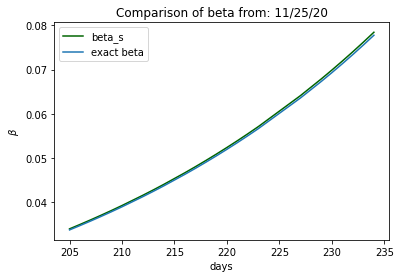}
    \caption{Comparison between the different betas from 11-25-2020}
    \label{lab15}
\end{figure}

\begin{figure}[H]
    \centering
    \includegraphics[scale=.55]{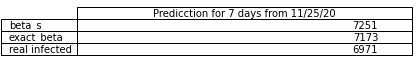}
    \caption{Prediction of infected people depending on $\beta$ and $\beta_S$}
    \label{lab16}
\end{figure}

\begin{figure}[H]
    \centering
    \includegraphics[scale=.5]{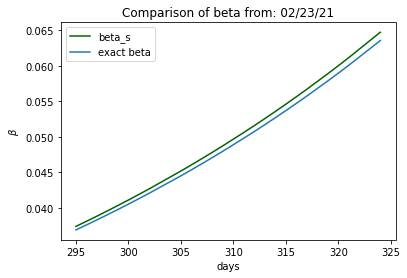}
    \caption{Comparison between the different betas from 02-23-2021}
    \label{lab17}
\end{figure}

\begin{figure}[H]
    \centering
    \includegraphics[scale=.7]{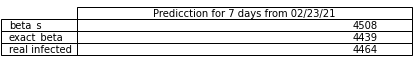}
    \caption{Prediction of infected people depending on $\beta$ and $\beta_S$}
    \label{lab18}
\end{figure}

\begin{figure}[H]
    \centering
    \includegraphics[scale=.5]{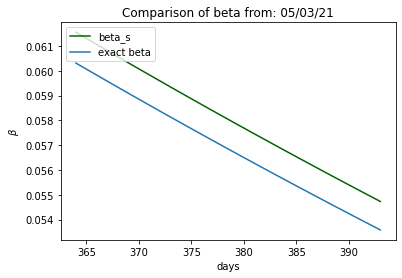}
    \caption{Comparison between the different betas from 05-03-2021}
    \label{lab19}
\end{figure}
\begin{figure}[H]
    \centering
    \includegraphics[scale=.5]{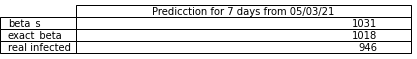}
    \caption{Prediction of infected people depending on $\beta$ and $\beta_S$}
    \label{lab20}
\end{figure}

As we can see in figure 1, figure 3 and figure 5 $\beta$s is a good approximation of $\beta_r$ in early stages on the pandemic as we mentioned in section II. But as time progresses, the difference between both increases.

\footnote{All predictions in this section were made assuming a detection rate $k = 0.10$.}

\input{Sections/6detection}

\input{Sections/7vaccination}


\section{Predictions}\label{sec 8}
\input{Sections/predicciones7dias}

\section{Conclusions and Outlook}

In this work we developed various methods to study the evolution of the COVID-19 pandemic. We concentrate in
the population evolution of Guanajuato state and the Mexican Republic. We consider the SIR model and modified versions of it. The main modified version is obtained
by defining the parameter $\beta_s=beta S$ which also casts the effect of the vaccination 
process. This occurs because the susceptible population $S$ can decrease as a consequence of
the vaccination and without incorporating a new population in the system,
the systematic fitting of $\beta_S$ can capture this feature.   In several cases,
we have found great agreement. Some of those extrapolations consider 21 days, but
others consider 30 days.

We explore as well the detection rate defined as the ratio between reported COVID recovered, infected and deceased populations  against the real ones. We have estimated this parameter to be of the order of $10\%$ in several cases. We perform a day-by-day fitting, in an interval of 21 days. In a given detection rate parameter interval, one chooses the rate daily,
which minimizes the difference between the predictions and the observed data. We then find an average between those values.  We also estimated this rate by means of a global fitting, be computed predictions
with different detection rates, choosing the one that minimizes the difference between 
observed data and real data.

The predictions computed have been performed with old data, and by taking actual points
to compare with. In a sort of checking prediction against actual developments. We finalize with up-to-date predictions of Mexico and Guanajuato. Additionally, we provide  a Github
program, where the reader can employ freely our work to check the predictions made here or to check  different cases.

\section{Acknowledgments}

We thank the Social Program Service of the Univesity of Guanajuato that allowed us to work on this project;
and the collaboration of the students: Gabriel Am\'ezquita, Oscar Esaul Cervantes, Juan Carlos God\'{\i}nez e Iv\'an Yebra. We thank the researchers: Argelia Bernal, Juan Barranco, Alejandro Cabo, Alma Gonz\'alez, Dami\'an Mayorga, Gustavo Niz y Luis Ure\~na for fruitful discussions. We thank the support of the project  CIIC 264/2022 UG and the Project CONACyT A1-S-37752.\\
All code used in this project can be found in https://github.com/JonathanHidalgoN/Covid-analysis.
\phantom{
\cite{ghosh2021modeling}\cite{honfo2020modeling}\cite{martinez2021caracterizacion}\cite{cordova2022adding}\cite{galban2021covid}\cite{lavielle2021predicting}\cite{gil2020estimating}\cite{cruz2021comportamiento}}

\section{Bibliography}
\bibliographystyle{plain}
\bibliography{Sections/biblioCOVID}

\input{Sections/9deathsregression}
\input{Sections/10kminimum}
\input{Sections/factor_lambda}
\input{Sections/Final_predicciones}

\end{document}

%% file: Sections/4betaS.tex
\section{Estimating the contagion rate via the $\beta_S$ model}\label{sec 4}

Given the SIR equations, it is possible to simplify the system of equations by fitting the $\beta(t) S(t)$ terms into a single function, $\beta_S(t)$. By doing this, the SIR system reduces to the following pair of equations:

\begin{equation}
    \dv{I}{t}=\beta_S I-\gamma_{eff}I
\end{equation}
\begin{equation}
    \dv{\Tilde{R}}{t}=\gamma_{eff}I
\end{equation}
The $\beta_S$ function is relevant because it captures the impact that the vaccination process has in the evolution of the pandemic without the need of introducing that information into the model. This happens
because the vaccinated population decreases the \textit{total} susceptible population (although it has been proved that vaccines don't necessarily limit the spread of the virus, they limit the severity of the infections; however, it can be hypothesized that this has an impact in the \textit{net} susceptible populations and in the \textit{net} transmission of the virus).

The fitting of $\beta_S$ is made by plotting $\gamma_{eff}+\dot{I}/I$ versus the time axis and fitting the points to an exponential function. We use as input the constant $\gamma_{eff}$ obtained in earlier sections.
Given that the numbers come from a daily report, $\dot{I}$ represent the new daily infected cases reported and $I$ represent the active cases of the day. By this, it can be noted that numerically, $\beta_{approx}$ and $\beta_S$ are identical but their meanings are completely different.


detected infections $I$, the $\beta_S$ terms need to be multiplied by a scaling factor. Let's say, if you go from the detected infections to the total infections $I\rightarrow I_{tot}=I/k$ and the inmune cases $\Tilde{R}\rightarrow \Tilde{R}_{tot}=\Tilde{R}/k$ (where $k$ is the detection rate), then $\beta_S\rightarrow \beta_{S,tot}=\lambda(k)\beta_S$, such that $\lambda(k)=S_{tot}/S$, where $S=1-(I+\Tilde{R})$ and $S_{tot}=1-(I_{tot}+\Tilde{R}_{tot})$ (the last expressions are valid using the units where the total population $N$ is set to 
$1$). Note that $\beta_S$ accounts for the change in the 
transformation between the \textit{reported} susceptible 
population and the real one. It is important to note that we use the term \textit{reported} when referring to infected because it 
is indirectly obtained by taking into consideration only the reported cases and comparing it to the total population, which is admittedly tricky and can be avoided -if wanted- by scaling the total population along the infected cases and the recovered ones. This is not what we do in this work for reasons that will become apparent when we estimate the detection rate. In other words, we construct quantities which are not all invariant by scaling the reported quantities to the real ones precisely to estimate the detection rate. In any case, the latter considerations cannot be applied to a model where precisely measured terms are considered (like the vaccination rate, for example), therefore, the simplification mentioned can only be used in the simpler models.

%% file: Sections/6detection.tex
\section{How to obtain the effective detection rate}\label{sec 6}
The \textbf{detection rate} is the proportion of the reported COVID-related statistics versus the real ones. Naturally, this number cannot be directly measured and, because of this, the estimation of this rate comes with a lot of uncertainty. This, of course, is a problem even if we don't take into consideration factors like the vaccination process.

Due to the \textit{vaccination independence} of the $\beta_S$ model, it is an useful tool in the estimation of the detection rate with only the most elemental factors: the reported infections, the recovered cases and the reported deaths. Nonetheless, it is important to say that the following method does not necessarily captures the real detection rate, but instead it gives an \textbf{effective detection rate}, which may come in handy only during an analysis based on the SIR model. However, in the following, we will limit to write \textit{detection rate} but it should be understood that it actually refers to the \textit{effective detection rate}.

The estimation of the detection rate goes as following:
\begin{enumerate}
    \item Choose an interval of time (in our case, we used the thirty most recent dairy registers on the database) such that even the most recent date plus the time window in which you make predictions is still in the dataset (in our case, such time window is 21 days). These set of registers from 52-21 days before the final date will be named \textbf{T-data}.
    \item On the interval $(0,1]$, which constitutes the possible values of the detection rate $k$, choose how many equidistant points you want to analyze (we used $100$ points, ranging from $0.01$ to $1.00$). The effective detection rate will arise from this set of values $k_{eff}\in (0,1]$. For simplicity, this set will be named the \textbf{k-interval}.
    \item For the oldest register in the T-data, make the predictions for the time window selected (21 days) for the different points on the k-interval. For each prediction, given a $k$ value, compare it with the real value using your preferred metric. For example, we used the sum of the relative errors between the prediction and the real point squared for the infected cases and the recovered ones. From all of those values, choose the value of $k$ which produces the least absolute difference between the predictions and the real value. This will be the characteristic $k$ value for this register.
    \item Repeat the last step for every register on the T-data.
    \item From the set of all characteristic $k$ values, we obtain the mean $k$ value and its respective standard deviation. This is the detection rate $k_{eff}$ and its associated estimation error $\Delta k_{eff}$.
\end{enumerate}

Note that it is possible to find some cases where the minimum $k$ on a given day is $1$. We ignored such cases for the calculation of $k_{eff}$ and $\Delta k_{eff}$, because we found that the reasonable values (by this, we mean the most frequent minimum $k$) are more close to $0.1$ than they are to $1$ (and, physically, this makes sense). As a matter of fact, those \textit{non-converging} values of $k$ \textbf{do} converge to a reasonable value if we increase the precision of the search. That is, studying smaller orders of magnitude for the decimals of $k$.

For México, analyzing the 5 months prior to 10-16-2021 with a precision up to 8 decimal places (using an optimized algorithm for searching around $k\sim0.1$), we obtain an effective detection rate of $0.015\pm0.022$. All the values of the time series range from $5.5560\times10^{-4}$ to $0.10000862$.

For Guanajuato, analyzing the 5 months prior to 10-18-2021 with a precision up to 8 decimal places (using said optimized algorithm for $k\sim0.1$), we obtain an effective detection rate of $0.027\pm0.019$. All the values of the time series range from $2.500\times10^{-4}$ to $0.0970$.

We also estimate the $k$ values by a different metric, which judges better the
global features. We calculate the difference between the population prediction of a given day
and the real population value, an sum the squares of the differences over the following 21 days to the T-data point. The populations are the infected and the active cases. We select the $k$ which minimizes this differences. There are ore details about it in \textbf{Section \ref{sec 10}}. For this method of estimation, for the same time period, we obtain $k_{global}=0.15$ for México and $k_{global}=0.04$ for Guanajuato.

\begin{figure}[H]
    \centering
    \includegraphics[scale=.55]{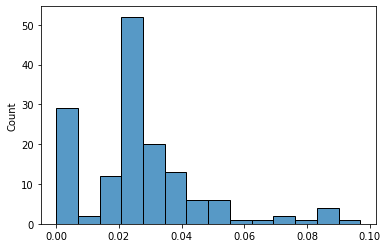}
    \caption{The histogram of the local detection rates for Guanajuato on the 5 months prior to 10-18-2021.}
    \label{fig:hist_gto}
\end{figure}

It should be noted that for Guanajuato, $k_{global}\sim k$. This property can be attributed to the distribution that the detection rates at different times show. In the \textbf{Figure \ref{fig:hist_gto}} it can be seen that the distribution is approximately normal. The normality of the distribution justifies the use of a mean value and a standard deviation to describe the overall behaviour of the sample. Hence, the global rate is similar to the mean rate.

\begin{figure}[H]
    \centering
    \includegraphics[scale=.55]{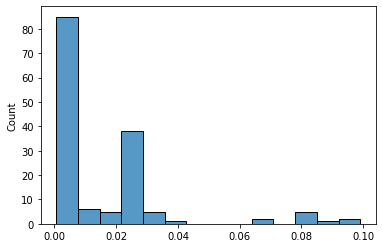}
    \caption{The histogram of the local detection rates for México on the 5 months prior to 10-16-2021.}
    \label{fig:hist_mex}
\end{figure}

For México, $k_{global}$ is not similar to $k$. The reason for this may be similar to that of Guanajuato, but in this case, it is because of the non-normality of the distribution shown in \textbf{Figure \ref{fig:hist_mex}}.
Point is, if someone found a method for constructing a normal distribution for the historical detection rates of México, then $k_{global}\sim k$.


%% file: Sections/7vaccination.tex
\section{Vaccination effects on the model}\label{sec 7}
The vaccination process intends to produce a decrease in the susceptibility of the population to the virus. By this, supposing that the vaccines are applied at a rate $\Tilde{\beta}$ and that they produce immediate immunity, then the SIR model can be modified as follows:
\begin{equation}
    \dv{S}{t}=-\beta SI-\Tilde{\beta}
\end{equation}
\begin{equation}
    \dv{I}{t}=\beta SI-\gamma_{eff}I
\end{equation}
\begin{equation}
    \dv{\Tilde{R}}{t}=\gamma_{eff}I
\end{equation}
\begin{equation}
    \dv{V}{t}=\Tilde{\beta}
\end{equation}
Where it can be easily noticed that the equations for the evolution of $I$ and $\Tilde{R}$ remain the same, while the $S$ equation adds a term considering the vaccination rate, which is defined on the fourth equation, where $V$ is precisely the vaccinated population. Notice that it is supposed that the set of people recovered and the vaccinated population are disjoint sets (which may be a valid supposition when $S>>\Tilde{R}$ because even if $S>>V$ or if $S\sim V$, then the contribution of $\Tilde{R}$ becomes either irrelevant together with the $V$ contribution or irrelevant when compared with the $V$ contribution).

It is important to mention that $V$ remains the same when $I$ and $\Tilde{R}$ are scaled by the detection rate because it is reasonable to assume that there is a strict control of the vaccines applied, such that all -or, in the worst case scenario, almost all- of the vaccinated people is properly taken into consideration when the statistics are reported. Therefore, this model isn't invariant when the aforementioned scaling is applied, just as discussed for the $\beta_S$ model.

\subsection{Comparison with $\beta_S$}
As we mentioned earlier, the $\beta_S$ model is interesting because it captures the effects of a lot of factors that may not be taken into consideration in all of the other discussed models. It is the case for the vaccination process. The $\Tilde{\beta}$ model for the vaccination works under the assumptions that the vaccine -whatever vaccine it is- produces an immediate perfect immunity and prevents completely the spread of the virus. Whoever, it is possibly to check if those assumptions are too much for it to make good predictions or not by, precisely, comparing a bunch of predictions using both models and then checking if those  predictions are adequate.

%% file: Sections/predicciones7dias.tex
In this section we show some predictions for Guanajuato and Mexico made with the different models and comparisons between the predictions and the difference with the real value. We employ the data shown in online by the sources \cite{datosGTO,datosMX}\\
\textbf{Guanajuato predictions.}\\
First predictions were made from day 05/25/21 to 06/17/21 blue line is the prediction made with $\beta_s$ model for 21 days.
\begin{figure}[H]
    \centering
    \includegraphics[scale=0.5]{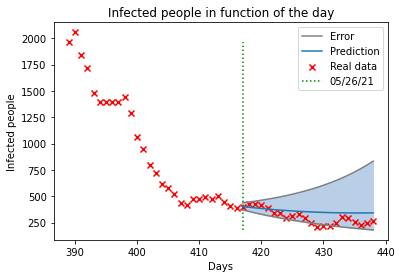}
    \caption{Red points show real infected people day to day, blue region represents the prediction for 21 days. }
    \label{lab1}
\end{figure}

\begin{figure}[H]
    \centering
    \includegraphics[scale=0.5]{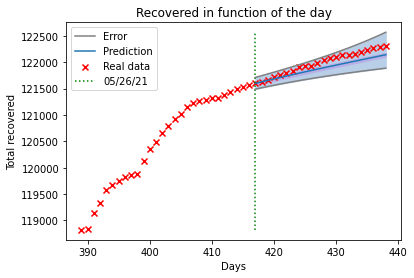}
    \caption{Red points show real recovery people day to day from, blue region represents the prediction for 21 days.}
    \label{lab2}
\end{figure}

\begin{figure}[H]
    \centering
    \includegraphics[scale=0.5]{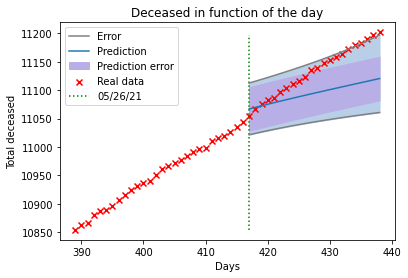}
    \caption{Red points show real deceased people day to day from, blue region represents the prediction for 21 days.}
    \label{lab3}
\end{figure}

In the next tables we can see the difference between real data and predictions with different models, where error was calculated with the difference in real data and the respective prediction.
\begin{figure}[H]
    \centering
    \includegraphics[scale=0.5]{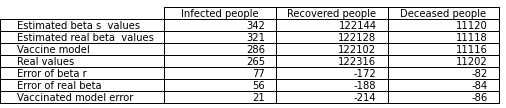}
    \caption{Comparison between the real data $\beta_s$, $\beta_r$ and vaccinate models for 21 days from 05/26/21.  }
    \label{lab4}
\end{figure}
 
\begin{figure}[H]
    \centering
    \includegraphics[scale=0.5]{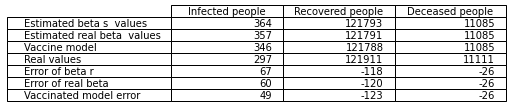}
    \caption{Comparison between the real data $\beta_s$, $\beta_r$ and vaccinate models for 7 days from 05/26/21 .  }
    \label{lab5}
\end{figure}

Here we show more predictions made from 4/18/21 to 5/12/21 for 21 days.
\begin{figure}[H]
    \centering
    \includegraphics[scale=0.5]{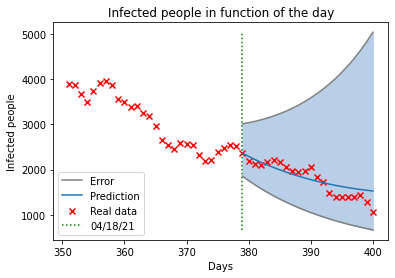}
    \caption{Red points show real infected people day to day, blue region represents the prediction for 21 days. }
    \label{lab6}
\end{figure}

\begin{figure}[H]
    \centering
    \includegraphics[scale=0.5]{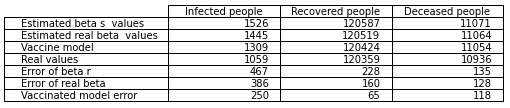}
    \caption{Comparison between the real data $\beta_s$, $\beta_r$ and vaccinate models for 21 days from 4/16/21.  }
    \label{lab7}
\end{figure}
\begin{figure}[H]
    \centering
    \includegraphics[scale=0.5]{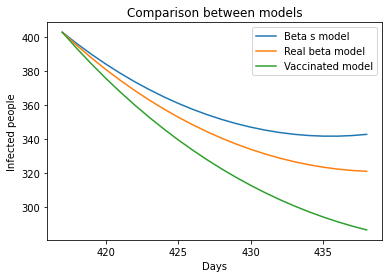}
    \caption{Comparison between the model predictions  $\beta_s$, $\beta_r$ and vaccinate models for 21 days from 5/26/21.  }
    \label{lab8}
\end{figure}

As we can see in tables in this section and figure 14 vaccinated model tends to report less infected people and it is a better approximation to real data.\\

\textbf{Mexico predictions.}\\
Predictions applying the $\beta_s$ model for the whole country with detection rate of 0.1. 

\begin{figure}[H]
    \centering
    \includegraphics[scale=0.5]{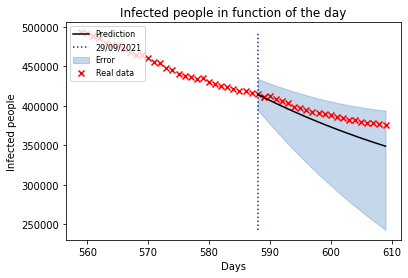}
    \caption{Mexico infected people predictions for 21 days from 29/09/2021, red dots represent real data and the black line predictions made with the $\beta_s$ model.}
    \label{lab9}
\end{figure}

\begin{figure}[H]
    \centering
    \includegraphics[scale=0.5]{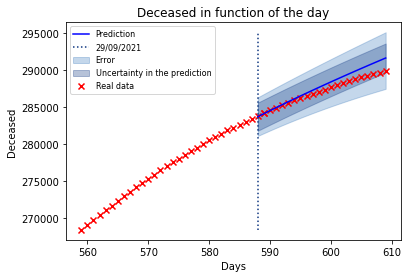}
    \caption{Mexico deceased people predictions for 21 days from 29/09/2021, red dots represent real data and the black line predictions made with the $\beta_s$ model.}
    \label{lab10}
\end{figure}

\begin{figure}[H]
    \centering
    \includegraphics[scale=0.5]{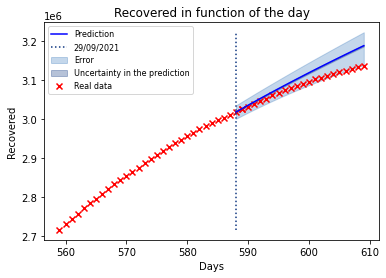}
    \caption{Mexico recovered people predictions for 21 days from 29/09/2021, red dots represent real data and the black line predictions made with the $\beta_s$ model.}
    \label{lab11}
\end{figure}
In general deceased and recover people predictions behave as expected, the prediction error encloses the real data.\\
More interesting is the infected people predictions as we can see in the next figure the predictions error do not enclose real data, even whit the big error region, the accuracy depends a lot on when we made the prediction, a possible explanation is that Mexico is a country with 128 million people and as we can see in figure 8 approximate the detection rate and other constants as the recovery or infection rate as if it is the same for the whole country is not optimal as Guanajuato predictions. 
\begin{figure}[H]
    \centering
    \caption{Mexico infected people predictions for 30 days from 29/06/2021, red dots represent real data, and the black line predictions made with the $\beta_s$ model.}
    \includegraphics[scale=0.5]{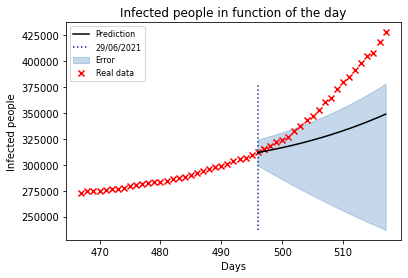}
    \label{lab12}
\end{figure}
To finish the predictions sections we present infected predictions for Guanajuato and México made with the $\beta_s$ model, in section \ref{sec 11} we present more predictions for Guanajuato and México.
\begin{figure}[H]
    \centering
    \includegraphics[scale=0.5]{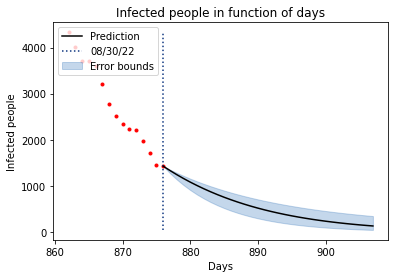}
    \caption{Guanajuato infected people from 15-06-2022 to 16-07-2022.}
    \label{lab13}
\end{figure}

\begin{figure}[H]
    \centering
    \includegraphics[scale=0.5]{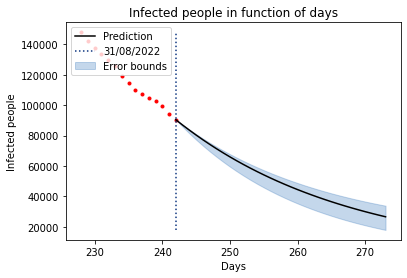}
    \caption{México deceased people from 14-06-2022 to 15-07-2022.}
    \label{lab14}
\end{figure}

%% file: Sections/9deathsregression.tex
\section{\textbf{Appendix}: Deaths linear regression}\label{sec 9}
It is necessary to remember that we can't directly make predictions about the number of recovered $R$ or deceased $D$ cases with any of the discussed models, that is because we used the \textit{immune} category, $\Tilde{R}=R+D$, which is the union of both sets. This means that it is necessary to make complementary adjustments of $R$ and $D$ both as functions of $\Tilde{R}$. We opted to use linear regressions, that is:
\begin{equation}
    R=\rho \Tilde{R}+b_\rho
\end{equation}
\begin{equation}
    D=\delta \Tilde{R}+b_\delta
\end{equation}
Where $\rho$ and $\delta$ are the slopes for the $R$ and $D$ fit and, likewise, $b_\rho$ and $b_\delta$ are the y-axis intersection for their respective straight line.

Intuitively, one should hope that both lines pass through the origin -and therefore, both $b_\rho$ and $b_\delta\approx 0$-, because that means that when there are $0$ immune people there are $0$ dead and recovered cases. However, this is usually not the case, because the y-axis intersections tend to \textit{not be that close to $0$}. There are a few of possible reasons.

The first possible explanation may be that the linear adjustment may not be a good fit and there may be a polynomial such that, for that expression, the intersection with the y-axis is, indeed, \textit{close to $0$}. It is possible to see if that's a possibility by doing just that for polynomials of various degrees and comparing the values of their y-axis intersections.

\begin{figure}[H]
    \centering
    \includegraphics[scale=.55]{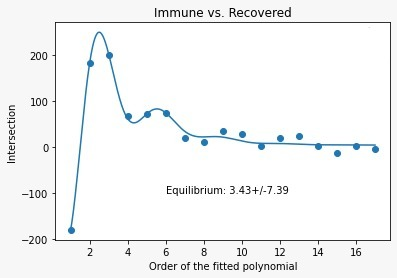}
    Optimal control of a SIR epidemic model with general incidence function and a time delays
    \caption{The intersection of the fit of $R(\Tilde{R})$ and the origin as a function of the degree of the polynomial fitted. This set of blue points is interpolated by a function (the blue curve) that behaves like a damped oscillator that tends to the equilibrium position shown. This shows that said intersection goes to zero with better polynomial approximations of real $R(\Tilde{R})$.}
    \label{fig:orderR}
\end{figure}

\begin{figure}[H]
    \centering
    \includegraphics[scale=.55]{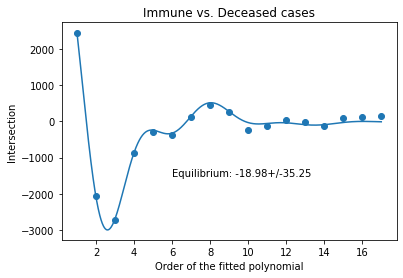}
    \caption{The intersection of the fit of $D(\Tilde{R})$ and the origin as a function of the degree of the polynomial fitted. This set of blue points is interpolated by a function (the blue curve) that behaves like a damped oscillator that tends to the equilibrium position shown. This shows that said intersection goes to zero with better polynomial approximations of real $R(\Tilde{R})$.}
    \label{fig:orderD}
\end{figure}

To check the first explanation, we ran an analysis running from April $4$th, $2020$ to the June $15$th, $2021$ (that is, the entire database until the day of the analysis). We $R$ as a function of $\Tilde{R}$ by fitting polynomials of different order and then checked the value of the intersection. The graph \ref{fig:orderR} shows the $y-axis$ intersection as a function of the degree of the fitted polynomial. Note that we fitted a function (using \textit{curve\_fit}, from \textbf{scipy}) to predict the behaviour of the intersection as the polynomial degree tends to infinity. We did the same procedure for $D$ as a function of $\Tilde{R}$, it is represented in the figure \ref{fig:orderD}.
The function fitted for \ref{fig:orderR} is (noting the intersection as $y_R$ and the degree as $n$):
\begin{equation*}
    y_R=-e^{-0.53n}\bigg(1,700cos(0.15n+1.5)+
\end{equation*}
\begin{equation}
    +440sin(1.9n-0.36)\bigg)+3.4
\end{equation}
With an error in the last term of $7.4$

The function fitted for \ref{fig:orderD} is (noting the intersection as $y_D$ and the degree as $n$):
\begin{equation*}
    y_D=e^{-0.33n}\bigg(5,300cos(0.66n+0.76)+
\end{equation*}
\begin{equation}
    +3,000sin(1.7n+0.31)\bigg)-19
\end{equation}
With an error in the last term of $35$.

By that analysis, note that only the last terms of the fit are relevant because as $n$ grows to infinity, the negative exponential will decrease until it is equal to zero. It can be noted that the terms that will survive aren't exactly zero, but it can be noted that zero is inside the interval of uncertainty. Therefore, we can see that the first explanation is plausible. 


The second explanation is complement to the first one. It may be that it is a consequence of the oscillations along the trend in both the recovered and deceased cases versus the total immune population. These oscillations may be interpreted as the back and forth between the population ignoring the safety measures when the things \textit{get better} and the population obeying such measures when things \textit{get worse}. Such perturbations may be the cause.

As a tangential mention, one could argue that there always can be lag and other error-inducing processes that happen between the hospital reports to the government and the reports that the government releases to the public, hence the variations along the linear fit.


%% file: Sections/10kminimum.tex
\section{\textbf{Appendix}: Another method for estimating the detection rate}\label{sec 10}

There is another possible way to estimate the detection rate. However, it isn't as efficient as the first method introduced and it doesn't lead to an easy estimation of the uncertainty of the rate.

The alternative estimation of the detection rate goes as following:
\begin{enumerate}
    \item On the interval $(0,1]$, which constitutes the possible values of the detection rate $k$, choose how many equidistant points you want to analyze (we used $100$ points, ranging from $0.01$ to $1.00$). The effective detection rate will arise from this set of values $k_{eff}\in (0,1]$. For simplicity, this set will be named the \textbf{k-interval}.
    \item Choose an interval of time (in our case, we used the thirty most recent daily registers on the dataset) such that even the most recent date plus the time window in which you make predictions is still in the dataset (in our case, such time window is 21 days). These set of registers from 52-21 days before the final date will be named \textbf{T-data}.
    \item For the first $k$ value in the k-interval, make the predictions for the time window selected (21 days) for every register in the T-data. For each prediction, given a register in the T-data, compare it with the real value using your preferred metric. For example, we used the sum of the relative errors between the prediction and the real point squared for the infected cases and the recovered ones. Sum over all of those error values.
    \item Repeat the last step for every $k$ value on the k-interval.
    \item Compare the sum of errors associated with every $k$ values. The $k$ value with the least global error is the estimated detection rate $k_{eff}$.
\end{enumerate}

Even if this method doesn't gives an estimation error, it is a good way to check if the effective detection rate obtained using the method presented in the \textbf{Section \ref{sec 6}} \textit{makes sense}.

%% file: Sections/factor_lambda.tex
\section{\textbf{Appendixi}: Lambda Factor and $\frac{dI}{dt}$}\label{sec 12}
\begin{figure}[H]
    \centering
    \includegraphics[scale=0.5]{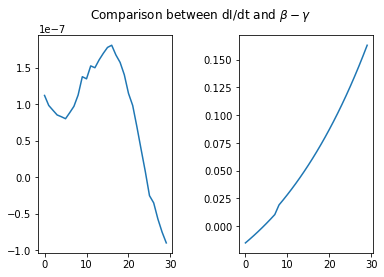}
    \caption{Comparison between dI/dt and $\beta_s-\gamma$ as we can see even when $\beta$-$\gamma$ > 0 dI/dt can be negative.}
    \label{dIdt}
\end{figure}

To make predictions we need data so we can compute quantities like infection rate or recovery rate but if we compare the predictions made with reality we observe that predictions always are below reality, this is well know because data recollected is just an approximation to real quantities that is why we use an the detection rate k but this have implications in SIR model, we refer to reported or observed quantities to data that is given by the government or different organizations and real quantities to an estimation of how much people is infected or has recovered in reality.\\
 Working with real quantities requires using the detection rate, we can transform between both if $I_o$ represents observed infected people then $I_r=\frac{I_o}{k}$ where $I_r$ are real infected people.
\\
If we ignore births $S_r+I_r+\tilde{R_r}=1$ where the subscript r refers to real quantities and the sum is one because we work with normalize population, $\tilde{R}$ represent the sum of recovered and dead people then we can write.
\begin{equation}\label{ec lambda factor 1}
    \frac{\tilde{R_o}}{k}+\frac{I_o}{k}+S_r=1
\end{equation}
From equation \ref{ec lambda factor 1}.
\begin{equation}\label{ec lambda factor 2}
    S_o=kS_{r}-\tilde{R}_o+1
\end{equation}
From equation \ref{ec lambda factor 2}.
\begin{equation}\label{lambda 4}
    \frac{S_r}{S_o} = \frac{1}{k}+\frac{\tilde{R}_r-\frac{1}{k} }{S_o}
\end{equation}
Working with equation (\ref{lambda 4}) and defining $\lambda=\frac{S_r}{S_o}$.
\begin{equation}\label{lambda final}
    \lambda=\frac{1-I_r-\tilde{R_r}}{1-k(I_r+\tilde{R_r})}
\end{equation}
With (\ref{lambda final}) the second equation in the SIR model its.
\begin{equation}\label{lambda factor 3}
    \frac{dI_r}{dt}=(\beta_s \lambda-\gamma_{eff})I
\end{equation}
Where $\beta_s=\beta S_o=\beta \lambda S_r$
\begin{figure}[H]
    \centering
    \includegraphics[scale=0.5]{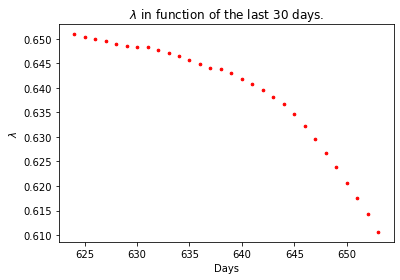}
    \caption{$\lambda$ factor in the function of time as we can see it is always decreasing with time.}
    \label{lambda graph}
\end{figure}
As we can see from definition of $\lambda$ its always decreasing with time and is bounded between 0 and 1 see figure ( \ref{lambda graph} ). This implies than even when $\beta_s$> $ \gamma_{eff}$ we can have $\frac{dI}{dt}$<0 see figure(\ref{dIdt}).

%% file: Sections/Final_predicciones.tex
\section{\textbf{Appendixi}: Final predictions for México and Guanajuato.}\label{sec 11}
In this section, we present final predictions for México and Guanajuato.
\begin{figure}[H]
    \centering
    \includegraphics[scale=0.5]{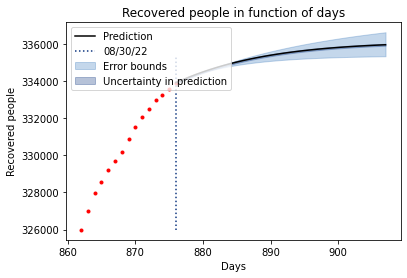}
    \caption{Guanajuato recovered people from 15-06-2022 to 16-07-2022.}
    \label{recpredfinal}
\end{figure}

\begin{figure}[H]
    \centering
    \includegraphics[scale=0.5]{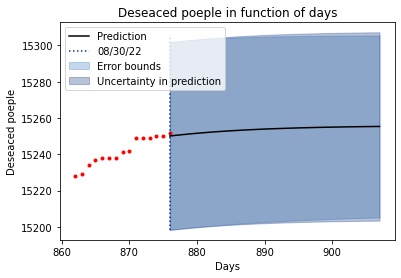}
    \caption{Guanajuato deceased people from 15-06-2022 to 16-07-2022..}
    \label{difuntospredfinal1}
\end{figure}

\begin{figure}[H]
    \centering
    \includegraphics[scale=0.5]{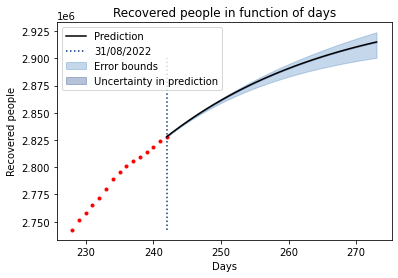}
    \caption{México recovered people from 14-06-2022 to 15-07-2022.}
    \label{recuperadospredfinal}
\end{figure}

\begin{figure}[H]
    \centering
    \includegraphics[scale=0.5]{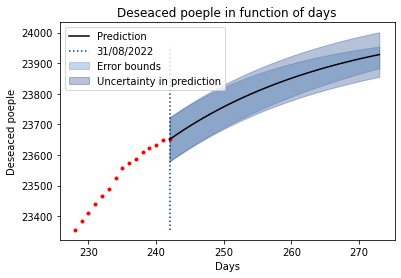}
    \caption{México deceased people from 14-06-2022 to 15-07-2022.}
    \label{difuntospredfinal}
\end{figure}